**Geology prediction based on operation data of TBM: comparison between deep neural network and statistical learning methods**

Maolin Shi[a], Xueguan Song[a,*] Wei Sun[a]

[a] School of Mechanical Engineering, Dalian University of Technology, Linggong Road, Dalian, China, 116024

[*]Corresponding author. Tel.: +86041184708410; fax: +86041184708410. E-mail address: sxg@dlut.edu.cn

**Abstract**

Tunnel boring machine (TBM) is a complex engineering system widely used for tunnel construction. In view of the complicated construction environments, it is necessary to predict geology conditions prior to excavation. In recent years, massive operation data of TBM has been recorded, and mining these data can provide important references and useful information for designers and operators of TBM. In this work, a geology prediction approach is proposed based on deep neural network and operation data. It can provide relatively accurate geology prediction results ahead of the tunnel face compared with the other prediction models based on statistical learning methods. The application case study on a tunnel in China shows that the proposed approach can accurately estimate the geological conditions prior to excavation, especially for the short range ahead of training data. This work can be regarded as a good complement to the geophysical prospecting approach during the construction of tunnels, and also highlights the applicability and potential of deep neural networks for other data mining tasks of TBMs.

## 1. Introduction

Tunnel boring machines (TBM) have been widely used for the tunnel construction because of their relatively high efficiency, safety and environmental friendliness compared to conventional blasting excavation [1-3]. During the tunneling process, a TBM excavates various geologies especially in the construction of metro tunnels, but the geological conditions are usually unknown prior to excavation [4-6]. Unknown geological conditions might bring huge damage to TBM, so it is necessary to develop methods to infer the geological conditions prior to excavation [7-10]. In recent years, a number of methods, including hard methods and soft methods, have been developed to infer the geological conditions prior to excavation in tunnel projects [1-2, 7-14]. Hard methods, including subsurface boring, pilot drilling and advanced geophysical prospecting, utilize in-site equipment to obtain geological information along the tunnel alignment [1-2]. However, they are usually not practical in real engineering practices since their high time and economic costs [4]. In contrast, soft methods, using statistical learning methods to estimate the geology conditions based on the geological information in some specific locations along the tunnel, are widely used to predict geology conditions in many tunnel projects. Alimoradi and Lau [5] used neural networks to predict the geological conditions based on the obtained geological information from the geological investigation report. Sun [11] utilized Kriging method to estimate the geological conditions and used the geology prediction results to help the load prediction of TBM. Sousa [12] used Bayesian networks to predict the geology conditions based on the performance of the tunnel boring machine. Miranda [13] used Bayesian updating and transition probability calculation to estimate the state probability of ground conditions along the tunnel alignment. Felletti and Beretta [14] used markov process approach to estimate the geology conditions based on the geological information revealed in some specific locations. Guan [4] improved their work, which can update the transition probability matrix dynamically along the tunnel. However, the geological information used in these statistical learning methods is only from limited specific locations along the tunnel, but not includes the whole geological information along the tunnel. On the other hand, most statistical learning methods have their special statistical assumptions (such as Kriging method assumes that all the attributes follows gaussian process, markov process approach assumes that the data is stationary), but the geologies in tunnel is difficult to follow these assumptions [2]. With the advancement and development of cyber-physical systems and measurement techniques, massive operation data of TBM are obtained during the excavation process. These data record not only the operation information of TBM but also the geological information [11]. Thus, mining these data is very useful for the geology prediction. However, the relationship between operation data and geological conditions is very nonlinear. In this work, a new method with strong nonlinear learning ability, deep neural networks, is introduced to predict the geological condition based on the operation data of TBM prior to excavation.

Deep neural networks is a type of machine learning method originating from artificial neural network, has drawn a lot of academic and industrial interest in recent years [15]. It uses multiple-layer architectures/deep architecture to extract the inherent features in data from the lowest layer to the highest layer, thus it can discover huge amounts of structure features including the complex relationship in the data set. It has been applied with success in engineering regression/classification tasks such as pattern recognition, image recognition, object detection, fault diagnosis and so on [16-20]. Google developed an image recognizer based on a nine-layered neural network and achieved the highest recognition rate in the international Imagenet Large Scale Visual Recognition Challenge competition in 2012 [16]. Tello [21] used deep neural networks to locate the root causes of failure in a semiconductor fabrication process, and demonstrated it achieved a better overall performance compared with traditional methods. Han [22] developed a geospatial object detection framework using a deep Boltzmann machine to assist the automatic interpretation of the optical remote sensing images. Tamilselvan [23] used deep belief networks for the health diagnosis of aircraft engine and electric power transformer, and AlThobiani [24] used it for the fault diagnosis of the valves in reciprocating compressors based on the vibration, pressure and current signals. Since the relationship between operation data and geological conditions is complicated in nature, deep neural networks can learn the complex relationship between the operation data and the corresponding geologies without any statistical assumptions, which has good performance for geology prediction.

In this work, a deep neural networks-based geology prediction approach is proposed aiming at predicting the geological conditions prior to excavation based on the operation data of TBM. To the best of the authors' knowledge, it is the first time that the deep neural networks is used to predict geological conditions for TBM. In addition, it demonstrates that the proposed geology prediction approach for has competitive performance compared with most soft geology prediction methods based on statistical learning methods. The rest of this paper is organized as follows. Section 2 presents the details of the proposed approach. Section 3 presents the geology prediction results on a tunnel constructed by TBM in China and their comparison with other geology prediction models based on statistical learning methods. Concluding remarks are described in Section 4.

## 2. Deep neural networks-based geology prediction approach

### 2.1 Deep neural networks

Deep neural networks originates from artificial neural networks. In 1989, Nielsen [25] proved the universal expressive power of three-layer nets through bumps and Fourier ideas. The proof indicates that any continuous functions from input to output can be implemented in a three-layer net, give sufficient number of hidden units and proper nonlinearities in activation function and weights. However, due to the lack of proper training algorithms, artificial neural network attracts less attention than the other statistical learning methods such as gaussian process and support vector machine until Hinton proposed deep learning in 2006 [16]. Deep learning involves a class of methods which try to hierarchically learn deep features of input data with very deep neural networks, typically deeper than three layers. It uses multiple-layer architecture/deep architecture to extract the inherent features in data from the lowest layer to the layer level. Thus, it can discover huge amounts of structure features including the complex relationship in the data set. According to some recent papers, it can give a better approximation to nonlinear functions than traditional statistical learning methods [26-28]. In this work, a deep neural networks is used to predict the geological conditions prior to excavation based on the operation data of TBM. A brief description of DNN and its training method used in this work is given as follows.

Artificial neural networks (ANN) have been developed as generalizations of mathematical models of biological nervous systems [16]. The basic element of ANN is artificial neuron (Fig. 1). For each artificial neuron, its output is computed as the weighted sum of the inputs, transformed by an activation function $f(\cdot)$ as follows.

$$Output = f(\sum_{i=1}^{m} \omega_i x_i + b) \tag{1}$$

where $\omega_i$ is the weight. The common activation functions include sigmoid function, tenhyperbolic function, softplus function and ReLu function as follows.

$$f_{sig} = 1/(1 + \mathrm{e}^{-\mathrm{x}}) \tag{2}$$

$$f_{tanh} = (e^x - e^{-x})/(e^x + e^{-x}) \tag{3}$$

$$f_{softplus} = log(1 + e^s) \tag{4}$$

$$f_{ReLu} = max(0, s) \tag{5}$$

Thus, the artificial neuron obtained the nonlinear capability by the help of the activation function. The learning capability of an artificial neuron is achieved by adjusting the weights in accordance to the chosen learning algorithm.

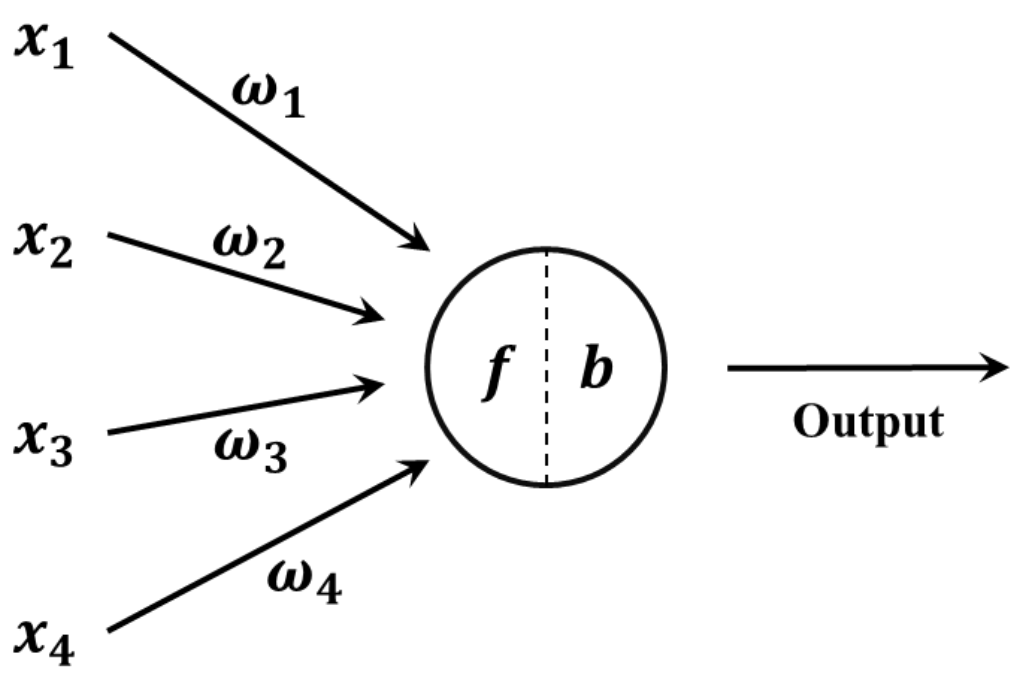

Figure 1 Artificial neuron

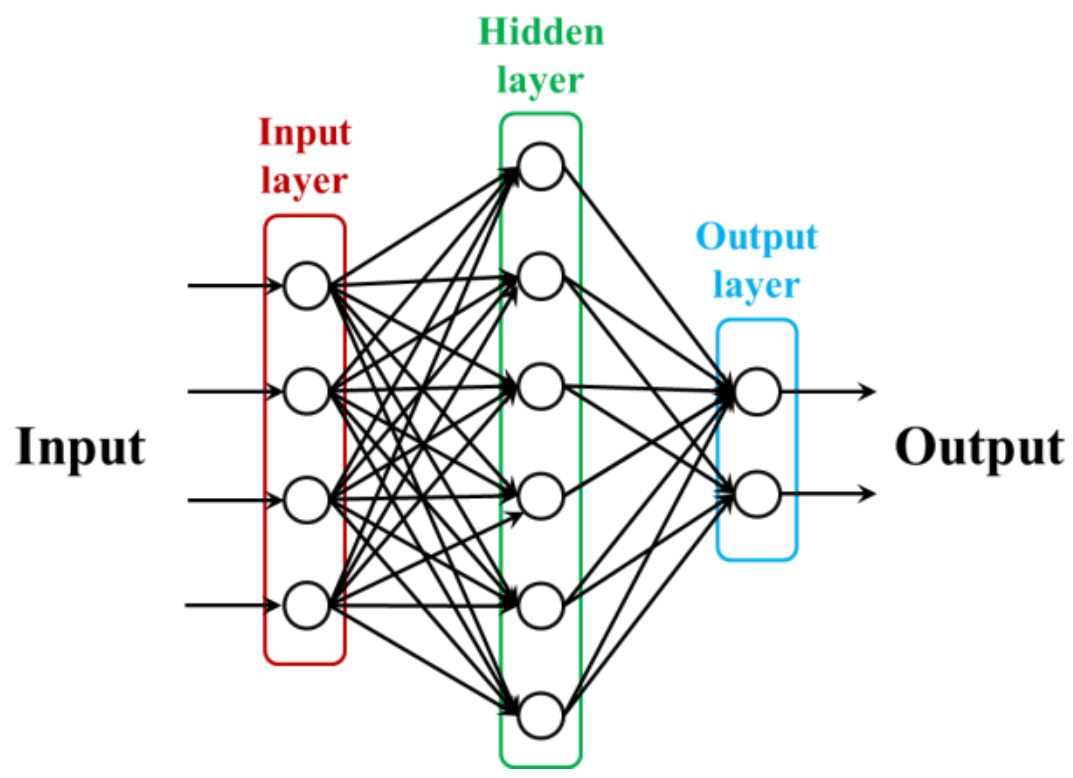

Figure 2 Artificial neural networks

Traditional artificial neural networks consists of three types of neuron layers: input, hidden, and output layers as shown in Fig. 2. The commonly used neural networks is feed-forward networks, which the data flow is from input to output, strictly in a feed-forward direction. There are several other neural network architectures such as Elman network and recurrent networks, and more details can be referred to Ref. [16] for an extensive overview of the different neural network architectures and learning algorithms.

The following discussion about deep learning and the proposed geology prediction approach in this work are both based on feed-forward networks (Fig. 3). From Figure 2 and 3, it can be found that the main difference between DNN and ANN is that deep neural networks use multiple-layer architecture by adding more layers into hidden layers. Thus, DNN obtains stronger nonlinear learning capability and is able to extract more inherent features in data from the lowest layer to the layer level. In this work, a DNN is used to predict the geology conditions based on the operation data of TBM prior to excavation.

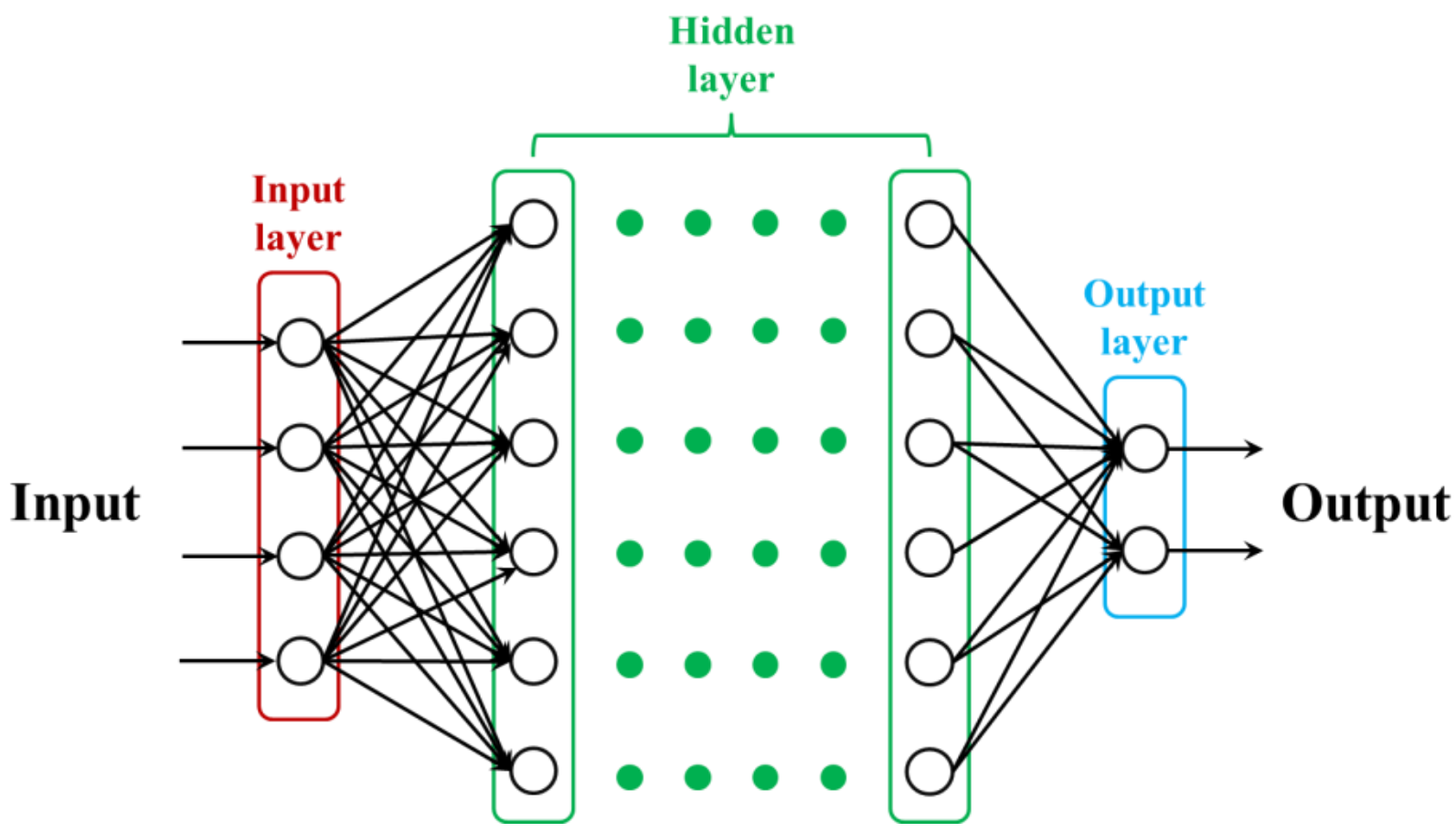

Figure 3 Deep neural networks

2.2 DNN-based geology prediction approach

With the advancement and development of cyber-physical systems and measurement techniques, massive operation data of TBMs are obtained during the excavation process. The operation data record the operation status of TBM and geology information, simultaneously. In this work, the operation data are used to predict geologies prior to excavation as shown in Figure 4.

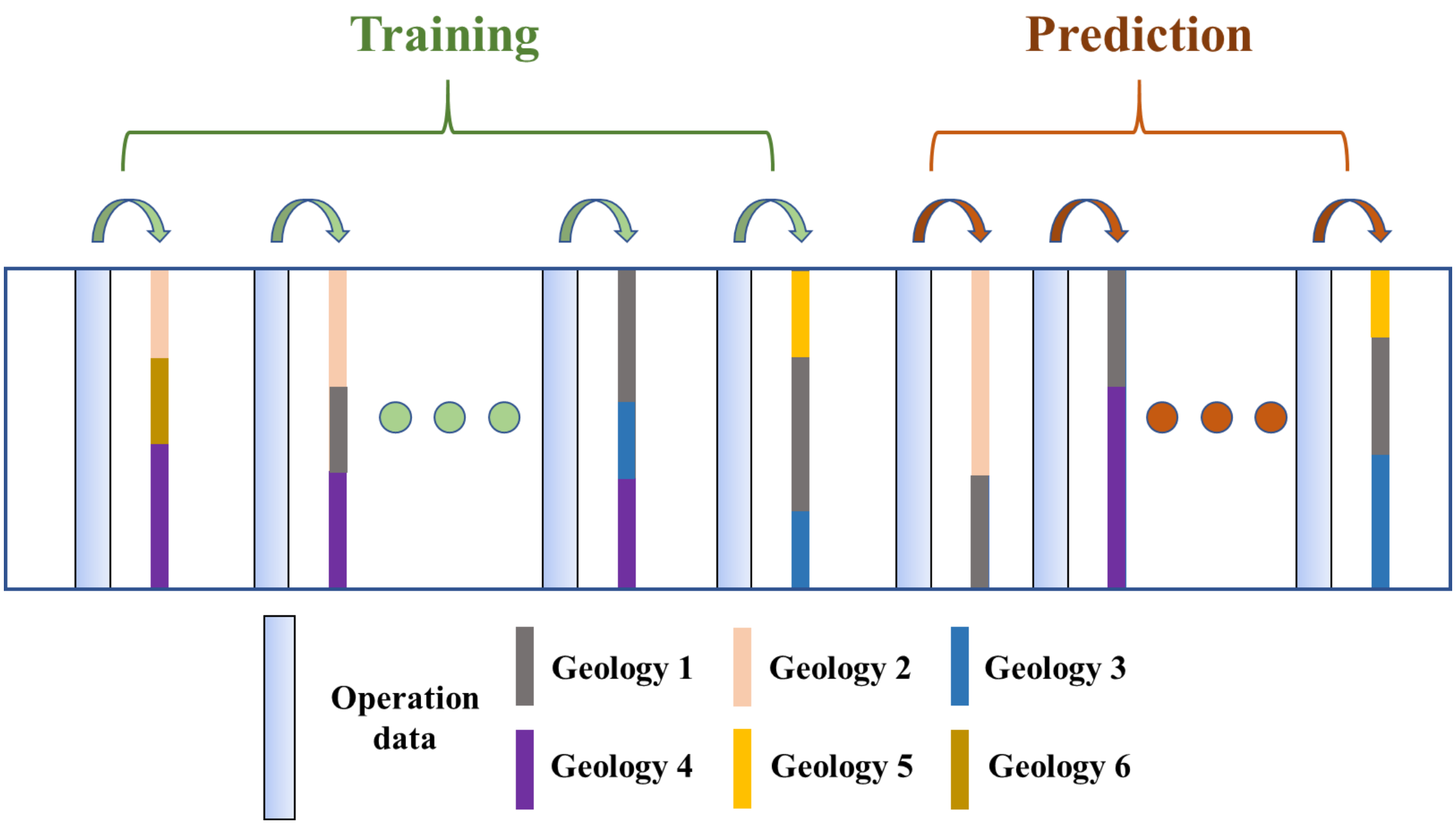

Figure 4 Geology prediction based on operation data of TBM

For each tunnel project, the strum is classified as several layers according to the geological investigation report before tunneling [1-2]. The proposed geology prediction approach is to predict one layer appears on the excavation face or not prior to excavation and to provide useful information to the operators and constructors of TBMs. In this work, operation data 1.0 meter before excavation face is used to predict the geologies appearing on the excavation face. For each geology, one DNN is built to predict it appears or not based on operation data. Thus, if there are *m* geologies exist in the tunnel, *m* DNNs are built. The DNNs used in this paper are built as follows.

### 2.2.1 Loss function

Denote $\widehat{y_{i,j}}$ as the output of the DNN-based predictors for the operation datum $\boldsymbol{x}$ and a special loss function categorical cross entropy [16] is used to measure the difference between the real output $y_{i,j}$ and the prediction output $\widehat{y_{i,j}}$.

$$L = \sum_{i=1}^{n} \sum_{j=1}^{2} -y_{i,j} log(\widehat{y_{i,j}}) \quad (6)$$

where $y_{i,j}$ is the real output, and $\widehat{y_{i,j}}$ is the predictions.

### 2.2.2 Training method

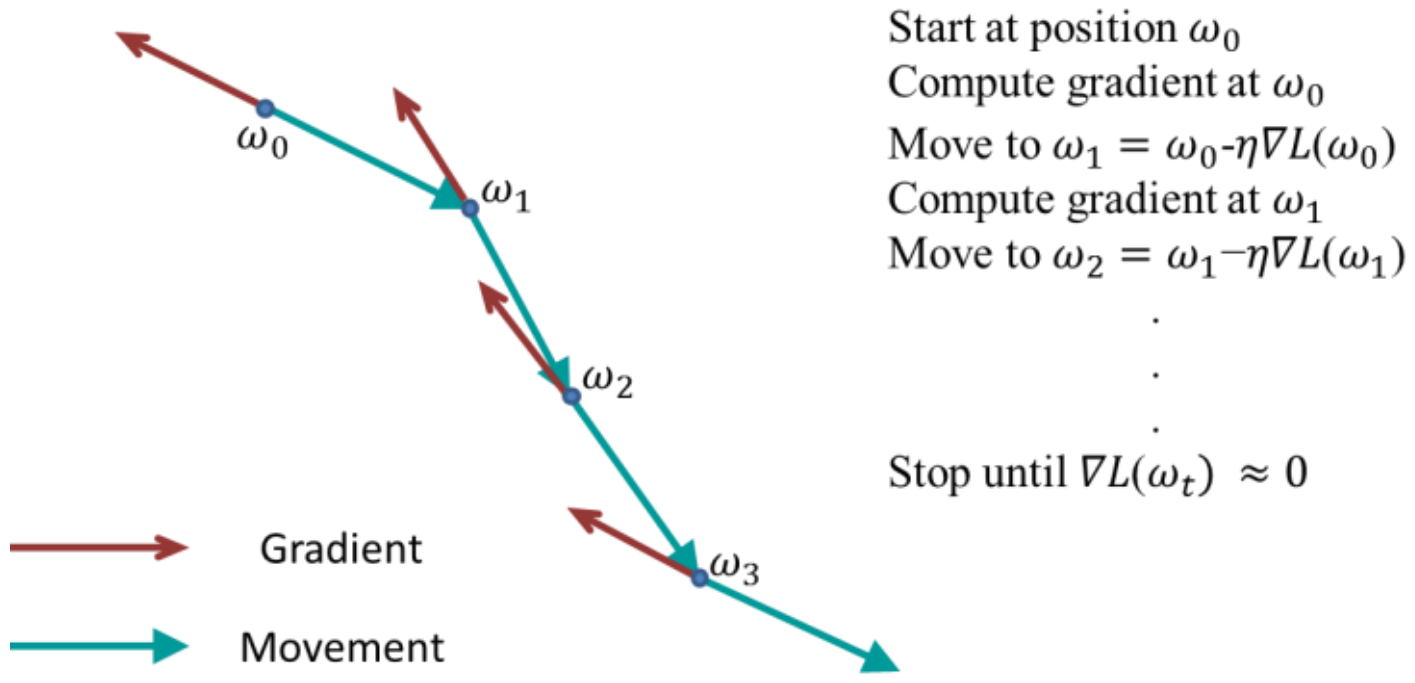

Figure 5 Gradient descent

The training of ANN is generally dependent on the gradient method as shown in Figure 5. In order to get better performance, a special gradient descent-base optimization method RMSpop [16] is used to minimize the loss function $L(\omega)$ which one can achieve the network training and then obtain the appropriate weights $\boldsymbol{\omega}$ for the resulting predictor. Compared with the traditional gradient descent optimization method with a constant learning rate $\eta$, RMSpop has a special learning rate as shown in Figure 6.

$$\omega_0$$

$$\omega_1 = \omega_1 - \frac{\eta}{\sigma_0} \nabla_0 \quad \sigma_0 = \nabla_0$$

$$\omega_2 = \omega_1 - \frac{\eta}{\sigma_1} \nabla_1 \quad \sigma_1 = \sqrt{\alpha(\sigma_0)^2 + (1-\alpha)(\nabla_1)^2}$$

$$\omega_3 = \omega_2 - \frac{\eta}{\sigma_2} \nabla_2 \quad \sigma_2 = \sqrt{\alpha(\sigma_1)^2 + (1-\alpha)(\nabla_2)^2}$$

$$\vdots$$

$$\omega_{t+1} = \omega_t - \frac{\eta}{\sigma_t} \nabla_t \quad \sigma_t = \sqrt{\alpha(\sigma_{t-1})^2 + (1-\alpha)(\nabla_t)^2}$$

Figure 6 RMSpop

The desired weights $\boldsymbol{\omega}$ can be obtained until the iteration rules such as maximum iterations reach.

### 2.2.3 Dropout

Because a fully connected layer occupies most of the parameters, it is prone to over-fitting. One important method to reduce over-fitting is dropout [30]. At each training stage, individual nodes are either "dropped out" of the networks with probability *p* or kept with probability 1-*p*, so that a reduced network is left; incoming and outgoing to a dropped-out node are also removed. Only the reduced network is trained on the data in that stage. The removed nodes are then reinserted into the network with their original weights. In the training stages, the probability that a hidden node will be dropped is usually 0.1~0.5. In this work, the probability of hidden nodes and input nodes is set at 0.2. Finally, the structure of the deep neural network used in this paper is as shown in Fig. 7.

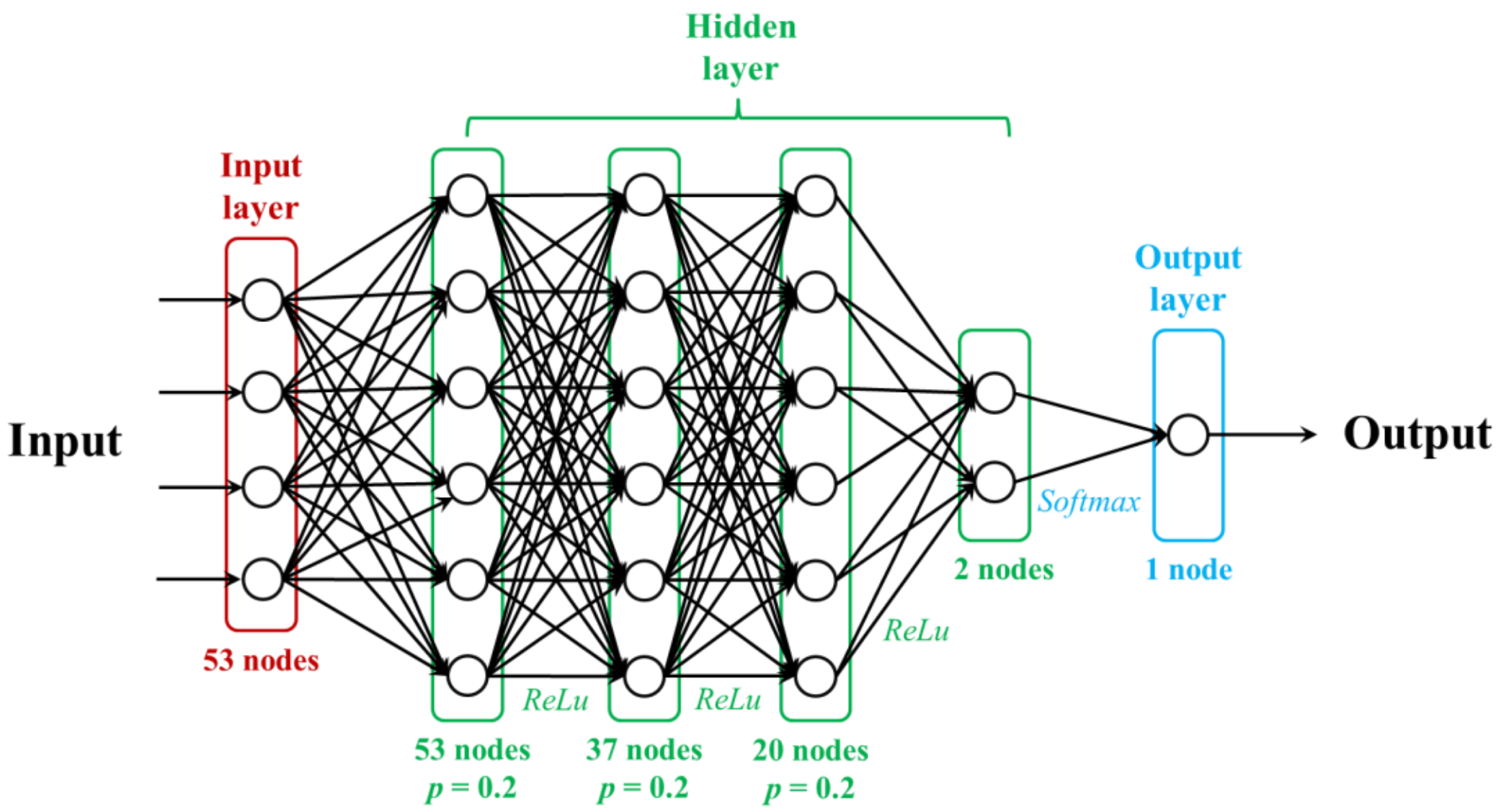

Figure 7.

## 2.3 Statistical learning methods

In recent decades, statistical learning methods are widely used for regression/classification tasks. To compare the performance of DNN with statistical learning methods, four popular statistical learning methods, logistic regression (LR) [30], naive Bayes classifiers (NBC) [31], random forest (RF) [32] and k-nearest neighbor (KNN) [33] are used to predict the geology based on the operation as well in this paper.

## 2.4 Measurement Indexes

Confusion matrix [34] is used and defined as follows:

Table 2 Confusion matrix

| | | True conditions | |
|---|---|---|---|
| | | Condition positive (*P*) | Condition negative (*N*) |
| Predicted condition | Condition positive (*P*) | True positive (*TP*) | False positive (*FP*) |
| | Condition negative (*N*) | False negative (*FN*) | True negative (*TN*) |

Condition positive (*P*): the number of real positive cases in the data. In this paper, it means the geology appears. Condition negative (*N*): the number of real negative cases in the data. In this paper, it means the geology does not appear. Based on the confusion matrix, Accuracy (*AC*), Matthews correlation coefficient (*MCC*) and Bookmaker Informedness (*BMI*) are used and defined as follows:

$$AC = \frac{TP+TN}{P+N} \times 100\% \tag{7}$$

$$MCC = \frac{TP\times TN - FP\times FN}{\sqrt{(TP+FP)(TP+FN)(TN+FP)(TN+FN)}} \tag{8}$$

$$BMI = \frac{TP}{P} + \frac{TN}{N} - 1 \tag{9}$$

The higher *AC* and *BMI* and the closer *MCC* to 1, the better the performance.

## 3. Engineering application

In this section, we provide the experimental results of the proposed approach for a real TBM. All experiments were processed by using Keras in a computer with Intel Core i7 CPU at 3.40 GHz, 16GB RAM and a NVIDIA GT1050T GPU.

### 3.1 Preject review

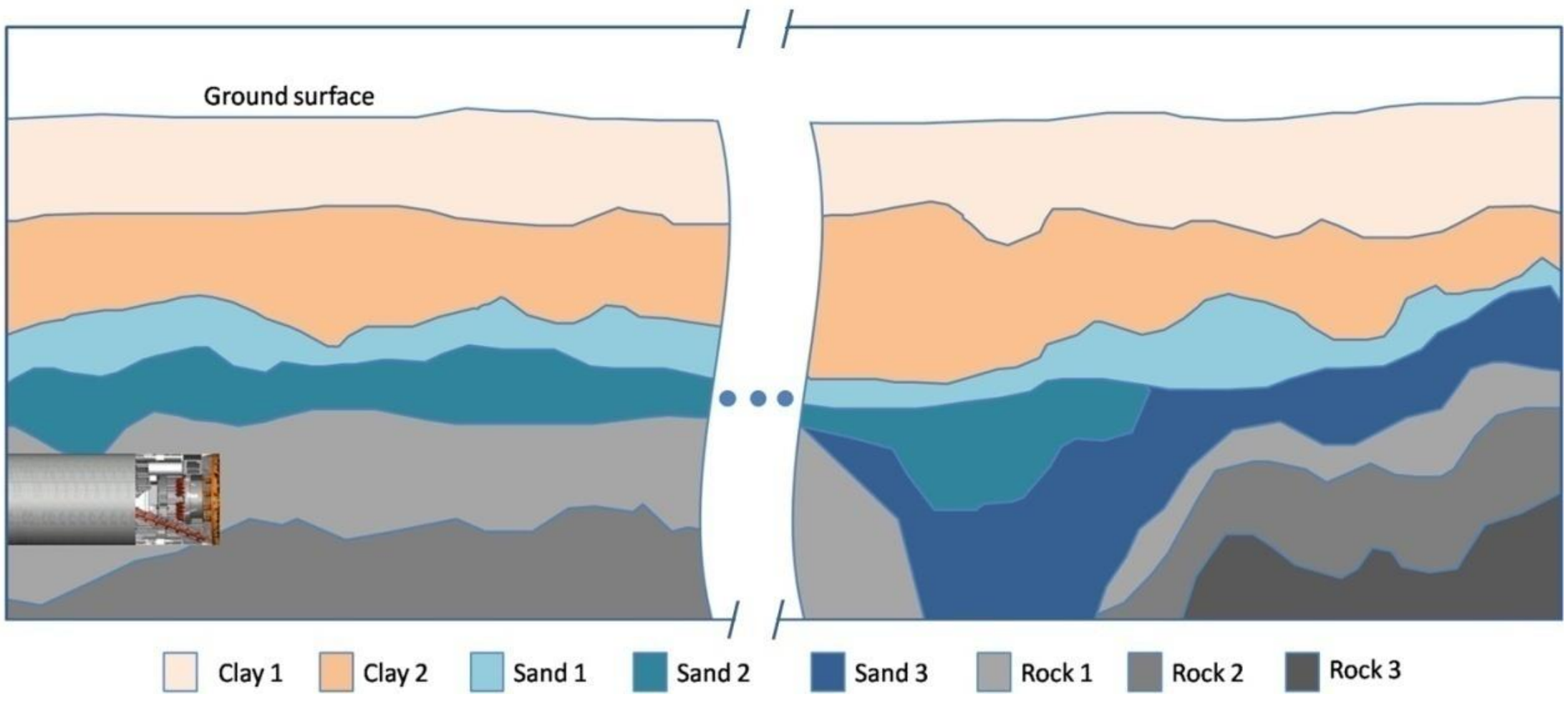

Figure 8 Longitudinal geological profile of the tunnel

The TBM operation data used here belong to a tunnel in China which has a length of 2000 m and a diameter of 6.4 m. A schematic illustration of the tunnel is provided in Fig 20. The ground surface elevation ranges from 0.2~5.8 m, and the depth of the tunnel floor from the ground surface ranges from 11.8~25.4 m. From the ground surface to the tunnel floor, various geological layers, such as clay, sand and rock, are unevenly distributed. Some of the geological characteristics of these layers are described in Appendix A. To excavate the tunnel, an earth pressure balance (EPB) shield TBM was used. This system consists of a cutterhead, chamber, screw conveyor, tail skin and other auxiliary subsystems. The TBM has a diameter of 6.2 m and a total mass of over 500,000 kg, and the cutterhead features an opening percentage of 30% and 120 cutters.

3.2 Geology description and preparation

The tunnel belongs to alluvial and coastal plain. The initial physiognomy is fishing pond and completely filled up in recent years. From the ground surface to the tunnel floor, there exist numerous soil, sand and rock stratums with uneven distribution. The engineering geological investigation report indicates the stratum appearing on the excavation face has eight geological layers as follows.

Layer 1: Dark gray clay. The surface elevation is -9.60~2.1m. The thickness is 1.60~12.40m with the average value 4.18m.

Layer 2: Quartz sand with little organic matter and more mud-silty clay. The thickness is 0.60~3.80m, generally 1.95m, and the surface elevation is -13.40~-4.61m.

Layer 3: Medium sand with 10~15% clay and less gravelly sand. The surface elevation is -2.10~-2.00m, and the thickness is 1.00~8.80m with mean value of 2.95m.

Layer 4: Brown yellow and hoary arenes. The surface elevation is -11.35 ~-3.45 m and the thickness is 0.50~4.50m, generally 2.18m.

Layer 5: Plastic sandy clay. The surface elevation is -22.19 ~ -2.54 m and the thickness is 0.7~19.0m with the mean value 5.09m.

Layer 6: Fully weathered rocks (hoary, brown red and brown yellow), extremely fractured with clear initial rock structure. The surface elevation is -25.01~4.72m.

Layer 7: Sandy weathered rocks (hoary and brown yellow) with loose structure, fractured, and its surface elevation is -28.57~5.72m.

Layer 8: Medium weathered rocks (brown yellow and brown red) with gneissose structure, little fractured. Its basic quantity of rock mass is IV. The surface elevation is - 53.83~-12.78m. The compressive strength is 14.6~38.8 MPa, generally 22.7MPa.

For each layer, if it appears on the excavation face, the corresponding geology is “1”; and if not, the geology is “0”. Thus, the geology data are obtained.

3.3 Operation data description and preparation

The operation data are composed of 53 attributes (for details, see Appendix A) that were continuously measured with a frequency of 1 Hz along the entire tunnel. 100s’ operation data before the location 1.0 m are used to predict the geologies of the corresponding location. In this paper, the geology information is obtained from 93 geology sampling locations along the tunnel, so 93 operation data sets containing a total of 9300*53 elements are obtained. These initial operation data sets inevitably have invalid values, and an index rate of change (*ROC*) is used to detect them []:

$$ROC_{x_t}^{+} = |log \frac{x_{t+1}}{x_t}| \quad (14)$$

$$ROC_{x_t}^{-} = |log \frac{x_{t-1}}{x_t}| \quad (15)$$

where $ROC_{x_t}^{+}$ is the *ROC* of attribute *x* between time *t* and *t*+1, and $ROC_{x_t}^{-}$ is the *ROC* of attribute *x* between time *t* and *t*-1. The criteria of $ROC_{x_t}^{+}$ and $ROC_{x_t}^{-}$ can be set to any value in terms of different applications. In this paper, both of them are set to 1.0; that is, the value of attribute *x* is deemed to be invalid only when both of them are larger than 1.0. The invalid values are handled as follows. For three or more consecutive invalid values, new values are estimated as follows.

$$y_j = \frac{z_m - z_0}{m} * j + z_0 \quad j \in (1,2,\ldots,m-1) \quad (16)$$

where $y_j$ is the *j-th* estimated value from the beginning of the set of consecutive invalid values, *m* is the number of consecutive invalid values, $z_m$ is the normal value next to the last invalid value, and $z_0$ is the normal value before the first invalid value. For one or two invalid values, the "persistence" method is adopted. This method involves using the normal value before the first invalid value to replace the following invalid values as follows:

$$y_n = z_0 \tag{17}$$

where $y_n$ is the invalid value. Then, the prepared operation data are combined with the geological data. Thus, 93 data sets are obtained. To validate the performance of the proposed geology prediction approach based on DNN, the former sequential 79 data sets are used as training data, and the latter 14 data sets are used as testing data. It is noted that the training data are sequentially but not randomly selected from the total data set (Fig. 4), which is more pratical for engineering practices [16, 34].

### 3.4 Results

In this section, the proposed geology prediction approach based on DNN is used to predict the geologies based on operation data. For each geological layer, one DNN is built and trained as discussed in Section 2. Eight DNN prediction models are built, and the prediction models based on different statistical learning models are built, respectively. The final results are presented in Figure 7. From this figure, it can be found that DNN achieves the best performance for seven geological layers and obtains competitive results for the other geological layer. The prediction accuracy of DNN is higher than 0.85 for geological layers 3, 5~8. The prediction accuracy for geological layers 2 and 4 are 0.7595 and 0.6788, respectively. It is noted that the prediction accuracy of geological layer 1 is relatively poor with 0.5263, and the prediction error are all false positive. The reason is that geological layer 1 is mostly located on the top area of the excavation face, and its ratio to total excavation face is relatively low compared with other geological layers. Thus, the operation data cannot include enough information for geological layer 1, thus the prediction performance of geological layer 1 is poor. Among statistical learning methods, NBC exhibits better performance than DNN on geological layer 2, and gets same prediction accuracy with DNN for geological layers 3 and 8; LR obtains same performance with DNN on geological layer 3; KNN obtains same performance with DNN on geological layer 8, and RF shows worse performance than DNN on all the geological layers. It can be found that DNN shows competitive geology prediction performance compared with statistical learning methods since its stronger nonlinear learning capability.

Since the training data and testing data are divided sequentially, the information contained in the training data is more associated with the geology conditions in short range of the tuunel face. The results of the closest 300 testing data to the training data are shown in Figure 8. It can be found that the proposed geology prediction approach based on DNN achieves competitive results. The prediction accuracy of DNN is 1.0000 for geological layers 1, 3, 5, 7~8. Geological layers 2 and 6 are 0.8900 and 0.7167, respectively. Thus, the proposed approach can accurately estimate the geological conditions prior to excavation, especially in short range of the tunnel face.

|  | | DNN P | DNN N | LR P | LR N | NBC P | NBC N | RF P | RF N | KNN P | KNN N |
|---|---|---|---|---|---|---|---|---|---|---|---|
| Layer 1 | P | 0 | 0 | 0 | 0 | 0 | 0 | 0 | 59 | 0 | 0 |
| | N | 900 | 1000 | 900 | 1000 | 900 | 1000 | 900 | 941 | 900 | 1000 |
| | AC | **0.5263** | | 0.5263 | | 0.5263 | | 0.4953 | | 0.5263 | |
| | MC | **NaN** | | NaN | | NaN | | -0.1698 | | NaN | |
| | BMI | **NaN** | | NaN | | NaN | | -0.059 | | NaN | |
| Layer 2 | P | 1155 | 312 | 1300 | 404 | 1300 | 404 | 0 | 19 | 1299 | 524 |
| | N | 145 | 288 | 0 | 196 | 0 | 196 | 1300 | 581 | 1 | 76 |
| | AC | **0.7595** | | 0.7874 | | 0.7874 | | 0.3058 | | 0.7237 | |
| | MC | **0.4083** | | 0.4992 | | 0.4992 | | -0.1479 | | 0.2968 | |
| | BMI | **0.3684** | | 0.3267 | | 0.3267 | | -0.0317 | | 0.1259 | |
| Layer 3 | P | 0 | 0 | 0 | 0 | 0 | 0 | 0 | 40 | 0 | 0 |
| | N | 0 | 1900 | 0 | 1900 | 0 | 1900 | 0 | 1860 | 42 | 1858 |
| | AC | **1.0000** | | 1.0000 | | 1.0000 | | 0.9789 | | 0.9779 | |
| | MC | **NaN** | | NaN | | NaN | | NaN | | NaN | |
| | BMI | **NaN** | | NaN | | NaN | | NaN | | NaN | |
| Layer 4 | P | 832 | 543 | 900 | 198 | 810 | 541 | 900 | 940 | 900 | 881 |
| | N | 68 | 457 | 0 | 802 | 90 | 459 | 0 | 60 | 0 | 119 |
| | AC | **0.6788** | | 0.5779 | | 0.6679 | | 0.5053 | | 0.5363 | |
| | MC | **0.4263** | | 0.3235 | | 0.3954 | | 0.1713 | | 0.2452 | |
| | BMI | **0.3819** | | 0.1980 | | 0.3590 | | 0.0600 | | 0.1190 | |
| Layer 5 | P | 1600 | 269 | 1600 | 300 | 1600 | 300 | 57 | 0 | 1600 | 289 |
| | N | 0 | 31 | 0 | 0 | 0 | 0 | 1543 | 300 | 0 | 11 |
| | AC | **0.8584** | | 0.8421 | | 0.8421 | | 0.1879 | | 0.8478 | |
| | MC | **0.2974** | | NaN | | NaN | | 0.0761 | | 0.1762 | |
| | BMI | **0.1033** | | 0 | | 0 | | 0.0356 | | 0.0367 | |
| Layer 6 | P | 15 | 70 | 0 | 102 | 200 | 1674 | 0 | 58 | 16 | 159 |
| | N | 185 | 1630 | 200 | 1598 | 0 | 26 | 200 | 1642 | 184 | 1541 |
| | AC | **0.8658** | | 0.8411 | | 0.1189 | | 0.8642 | | 0.8195 | |
| | MC | **0.0502** | | -0.0816 | | 0.0404 | | -0.0609 | | -0.0143 | |
| | BMI | **0.0338** | | -0.0600 | | 0.0152 | | -0.0341 | | -0.0135 | |
| Layer 7 | P | 90 | 0 | 75 | 16 | 75 | 16 | 0 | 32 | 0 | 0 |
| | N | 110 | 1700 | 125 | 1684 | 125 | 1684 | 200 | 1668 | 200 | 1700 |
| | AC | **0.9421** | | 0.9258 | | 0.9258 | | 0.8779 | | 0.8947 | |
| | MC | **0.6501** | | 0.5254 | | 0.5254 | | -0.0449 | | NaN | |
| | BMI | **0.4500** | | 0.3655 | | 0.3655 | | -0.0188 | | 0 | |
| Layer 8 | P | 0 | 0 | 0 | 1 | 0 | 0 | 0 | 38 | 0 | 0 |
| | N | 0 | 1900 | 0 | 1899 | 0 | 1900 | 0 | 1862 | 0 | 1900 |
| | AC | **1.000** | | 0.9995 | | 1.0000 | | 0.9800 | | 1.0000 | |
| | MC | **NaN** | | NaN | | NaN | | NaN | | NaN | |
| | BMI | **NaN** | | NaN | | NaN | | NaN | | NaN | |

Figure 9 Prediction results

| Layer 1 | DNN | | LR | | NBC | | RF | | KNN | |
|---|---|---|---|---|---|---|---|---|---|---|
| | *P* | *N* | *P* | *N* | *P* | *N* | *P* | *N* | *P* | *N* |
| *P* | 0 | 0 | 0 | 0 | 0 | 0 | 0 | 1 | 0 | 0 |
| *N* | 0 | 300 | 0 | 300 | 0 | 300 | 0 | 299 | 0 | 300 |
| *AC* | 1.0000 | | 1.0000 | | 1.0000 | | 0.9967 | | 1.0000 | |
| *MC* | NaN | | NaN | | NaN | | NaN | | NaN | |
| *BMI* | NaN | | NaN | | NaN | | NaN | | NaN | |

| Layer 2 | DNN | | LR | | NBC | | RF | | KNN | |
|---|---|---|---|---|---|---|---|---|---|---|
| | *P* | *N* | *P* | *N* | *P* | *N* | *P* | *N* | *P* | *N* |
| *P* | 176 | 9 | 200 | 100 | 200 | 100 | 0 | 0 | 199 | 100 |
| *N* | 24 | 91 | 0 | 0 | 0 | 0 | 200 | 100 | 1 | 0 |
| *AC* | 0.8900 | | 0.6667 | | 0.6667 | | 0.3333 | | 0.6633 | |
| *MC* | 0.7660 | | NaN | | NaN | | NaN | | -0.0408 | |
| *BMI* | 0.7900 | | 0 | | 0 | | 0 | | -0.0050 | |

| Layer 3 | DNN | | LR | | NBC | | RF | | KNN | |
|---|---|---|---|---|---|---|---|---|---|---|
| | *P* | *N* | *P* | *N* | *P* | *N* | *P* | *N* | *P* | *N* |
| *P* | 0 | 0 | 0 | 0 | 0 | 0 | 0 | 0 | 0 | 0 |
| *N* | 0 | 300 | 0 | 300 | 0 | 300 | 0 | 300 | 0 | 300 |
| *AC* | 1.0000 | | 1.0000 | | 1.0000 | | 1.0000 | | 1.0000 | |
| *MC* | NaN | | NaN | | NaN | | NaN | | NaN | |
| *BMI* | NaN | | NaN | | NaN | | NaN | | NaN | |

| Layer 4 | DNN | | LR | | NBC | | RF | | KNN | |
|---|---|---|---|---|---|---|---|---|---|---|
| | *P* | *N* | *P* | *N* | *P* | *N* | *P* | *N* | *P* | *N* |
| *P* | 0 | 169 | 0 | 300 | 0 | 164 | 0 | 300 | 0 | 300 |
| *N* | 0 | 131 | 0 | 0 | 0 | 136 | 0 | 0 | 0 | 0 |
| *AC* | 0.4367 | | 0.0000 | | 0.4533 | | 0.0000 | | 0.0000 | |
| *MC* | NaN | | NaN | | NaN | | NaN | | NaN | |
| *BMI* | NaN | | NaN | | NaN | | NaN | | NaN | |

| Layer 5 | DNN | | LR | | NBC | | RF | | KNN | |
|---|---|---|---|---|---|---|---|---|---|---|
| | *P* | *N* | *P* | *N* | *P* | *N* | *P* | *N* | *P* | *N* |
| *P* | 300 | 0 | 300 | 0 | 300 | 0 | 0 | 0 | 300 | 0 |
| *N* | 0 | 0 | 0 | 0 | 0 | 0 | 300 | 0 | 0 | 0 |
| *AC* | 1.0000 | | 1.0000 | | 1.0000 | | 0.0000 | | 1.0000 | |
| *MC* | NaN | | NaN | | NaN | | NaN | | NaN | |
| *BMI* | NaN | | NaN | | NaN | | NaN | | NaN | |

| Layer 6 | DNN | | LR | | NBC | | RF | | KNN | |
|---|---|---|---|---|---|---|---|---|---|---|
| | *P* | *N* | *P* | *N* | *P* | *N* | *P* | *N* | *P* | *N* |
| *P* | 15 | 0 | 0 | 1 | 100 | 200 | 0 | 0 | 16 | 100 |
| *N* | 85 | 200 | 100 | 199 | 0 | 0 | 100 | 200 | 84 | 100 |
| *AC* | 0.7167 | | 0.6633 | | 0.6666 | | 0.6666 | | 0.3867 | |
| *MC* | 0.3244 | | -0.0409 | | NaN | | NaN | | -0.3291 | |
| *BMI* | 0.1500 | | -0.0050 | | 0 | | 0 | | -0.3400 | |

| Layer 7 | DNN | | LR | | NBC | | RF | | KNN | |
|---|---|---|---|---|---|---|---|---|---|---|
| | *P* | *N* | *P* | *N* | *P* | *N* | *P* | *N* | *P* | *N* |
| *P* | 0 | 0 | 0 | 0 | 0 | 11 | 0 | 0 | 0 | 0 |
| *N* | 0 | 300 | 0 | 300 | 0 | 289 | 0 | 300 | 0 | 300 |
| *AC* | 1.0000 | | 1.0000 | | 0.9633 | | 1.0000 | | 1.0000 | |
| *MC* | NaN | | NaN | | NaN | | NaN | | NaN | |
| *BMI* | NaN | | NaN | | NaN | | NaN | | NaN | |

| Layer 8 | DNN | | LR | | NBC | | RF | | KNN | |
|---|---|---|---|---|---|---|---|---|---|---|
| | *P* | *N* | *P* | *N* | *P* | *N* | *P* | *N* | *P* | *N* |
| *P* | 0 | 0 | 0 | 0 | 0 | 0 | 0 | 0 | 0 | 0 |
| *N* | 0 | 300 | 0 | 300 | 0 | 300 | 0 | 300 | 0 | 300 |
| *AC* | 1.000 | | 1.0000 | | 1.0000 | | 1.0000 | | 1.0000 | |
| *MC* | NaN | | NaN | | NaN | | NaN | | NaN | |
| *BMI* | NaN | | NaN | | NaN | | NaN | | NaN | |

Figure 10 Prediction results

## 4. Conclusions

In this work, a geology prediction approach is proposed based on a five-layers deep neural networks and operation data. In the deep neural networks, categorical cross entropy is used as the loss function considering the unbalance of geology data, a special gradient descent-base optimization method RMSpop is used to minimize the loss function, and dropout is used to reduce over-fitting. The application case study on a tunnel in China shows that the proposed approach can accurately estimate the geological conditions prior to excavation compared with the other prediction models based on statistical learning methods LR, NBC, RF and KNN. This work can be regarded as a good complement to the geophysical prospecting approach during the construction of tunnels, and also highlights the applicability and potential of deep neural networks for other data mining tasks of TBMs.

## Acknowledgements

The research is supported by National Natural Science Foundation of China (Grant No. 51505061 and U1608256).

Appendix A:

Table 1 Attributes and abbreviations

| Abbreviation | Attribute | Abbreviation | Attribute |
|---|---|---|---|
| TOT | Temperature of oil tank (℃) | TGO | Temperature of gear oil (℃) |
| RC | Rotation speed of cutterhead (r/min) | FP | Propelling pressure (bar) |
| FPA | Pressure of A group of hydraulic cylinders (bar) | FPB | Pressure of B group of hydraulic cylinders (bar) |
| FPC | Pressure of C group of hydraulic cylinders (bar) | FPD | Pressure of D group of hydraulic cylinders (bar) |
| PEB | Pressure of equipment bridge (bar) | PA | Pressure of articulation system (bar) |
| PTSTRF | Pressure of tail skin system at top right front (bar) | PTSRF | Pressure of tail skin system at right front (bar) |
| PTSBRF | Pressure of tail skin system at bottom right front (bar) | PTSTLF | Pressure of tail skin system at top left front (bar) |
| PTSLF | Pressure of tail skin system at left font (bar) | PTSBLF | Pressure of tail skin system at bottom left front (bar) |
| PTSTRB | Pressure of tail skin system at top right back (bar) | PTSRB | Pressure of tail skin system at right front (bar) |
| PTSBRB | Pressure of tail skin system at bottom right back (bar) | PTSTLB | Pressure of tail skin system at top left back (bar) |
| PTSLB | Pressure of tail skin system at left back (bar) | PTSBLB | Pressure of tail skin system at bottom left back (bar) |
| RSC | Rotation speed of screw conveyor (r/min) | PSCP | Pressure of screw conveyor pump (bar) |
| TSC | Temperature of screw conveyor () | PSC | Pressure of screw conveyor (bar) |
| RA | Rolling angle (°) | PCTL | Pressure of chamber at top left (bar) |
| PCT | Pressure of chamber at top (bar) | PCBL | Pressure of chamber at bottom left (bar) |
| PCBR | Pressure of chamber at bottom right (bar) | PCTR | Pressure of chamber at top right (bar) |
| PB | Pressure of bentonite (bar) | GPTL | Grout pressure at top left (bar) |
| GPTR | Grout pressure at top right (bar) | GPBR | Grout pressure at bottom right (bar) |
| GPBL | Grout pressure at bottom left (bar) | BPSS | Bentonite pressure of shield shell (bar) |
| PSCF | Pressure of screw conveyor at front (bar) | PI | Penetration rate (mm/s) |
| T | Torque of cutterhead (kNm) | TSC | Torque of screw conveyor (Nm) |
| RBC | Rotation speed of belt conveyor (m/s) | SA | Displacement of A group of thrust cylinders (mm) |
| SB | Displacement of B group of thrust cylinders (mm) | SC | Displacement of C group of thrust cylinders (mm) |
| SD | Displacement of C group of thrust cylinders (mm) | SATR | Displacement of articulated system at top right (mm) |
| SABR | Displacement of articulated system at bottom right (mm) | SATL | Displacement of articulated system at top left (mm) |
| SABL | Displacement of articulated system at bottom left (mm) | F | Thrust of cutterhead (kN) |
| PIA | Pitch angle (°) | | |